\documentstyle[prd,aps,preprint,tighten,epsfig]{revtex}

\begin{document}

\draft

\title{The Majorana Neutrino Mass Matrix with One Texture Zero
and One Vanishing Eigenvalue}
\author{{\bf Zhi-zhong Xing}}
\address{CCAST (World Laboratory), P.O. Box 8730, Beijing 100080, China \\
and Institute of High Energy Physics, Chinese Academy of Sciences, \\
P.O. Box 918 (4), Beijing 100039, China 
\footnote{Mailing address} \\
({\it Electronic address: xingzz@mail.ihep.ac.cn}) } 
\maketitle

\begin{abstract}
Possible patterns of the Majorana neutrino mass matrix $M$ with
one texture zero and one vanishing eigenvalue are classified 
and discussed. We find that three one-zero textures of $M$ with
$m_1 =0$ and four one-zero textures of $M$ with $m_3 =0$ are 
compatible with current neutrino oscillation data. The implications
of these phenomenological ans$\rm\ddot{a}$tze on the neutrino mass
spectrum and the neutrinoless double beta decay are also explored 
in some detail.
\end{abstract}


\newpage

The recent SK \cite{SK}, SNO \cite{SNO}, KamLAND \cite{KM} and 
K2K \cite{K2K} neutrino oscillation experiments have 
provided us with very convincing evidence that neutrinos are massive 
and lepton flavors are mixed. At low energy scales, the phenomenology
of neutrino masses and lepton flavor mixing can be formulated by the
effective neutrino mass matrix $M$ in the flavor basis where the
charged lepton mass matrix is diagonal. Note that $M$ totally involves
nine real parameters: three neutrino masses ($m_1, m_2, m_3$),
three flavor mixing angles ($\theta_x, \theta_y, \theta_z$), 
one Dirac-type CP-violating phase ($\delta$) and two Majorana-type
CP-violating phases ($\rho$ and $\sigma$). Although we have achieved 
some quantitative knowledge on two neutrino mass-squared differences 
and three flavor mixing angles from current neutrino oscillation 
experiments, much more effort is needed to determine these parameters 
to a high degree of accuracy. The most challenging task is to pin down 
the absolute neutrino mass scale and the CP-violating phases. The
latter are entirely unrestricted by today's experimental data.

However, the presently conceivable sets of feasible neutrino 
experiments are only possible to determine seven of the nine 
independent parameters of $M$ \cite{FGM}. This observation implies 
that some phenomenological hypotheses have to be made for the texture
of $M$, in order to establish testable relations between the
experimental quantities and the parameters of $M$. 
In Ref. \cite{FGM}, the patterns of $M$ with two texture zeros
have been classified and discussed. In Ref. \cite{Branco}, 
particular attention has been paid to the possibility that $M$ may
have one vanishing eigenvalue (i.e., $m_1=0$ or $m_3=0$). Both
approaches allow us to determine or constrain the values of neutrino 
masses and Majorana phases in terms of the relevant experimental 
observables.

The purpose of this paper is to consider another simple but 
interesting approach for the study of neutrino masses and lepton
flavor mixing at low energy scales. 
We assume that the Majorana neutrino mass matrix
$M$ possesses one texture zero and one vanishing eigenvalue.  
Then we find that three one-zero patterns of $M$ with $m_1 =0$
(normal hierarchy) and four one-zero patterns of $M$ with 
$m_3 =0$ (inverted hierarchy) are compatible with current neutrino
oscillation data. The implications of these phenomenological
ans$\rm\ddot{a}$tze on the neutrino mass spectrum and the 
neutrinoless double beta decay are also discussed in some detail.

In the flavor basis where the charged lepton mass matrix is
diagonal, the neutrino mass matrix $M$ can be written as
\begin{equation}
M \; =\; V \left ( \matrix{
m_1 & 0 & 0 \cr
0 & m_2 & 0 \cr
0 & 0 & m_3 \cr} \right ) V^T \; ,
\end{equation}
where the lepton flavor mixing matrix $V$ can be parametrized 
as \cite{FX01}
\begin{equation}
V \; = \; \left ( \matrix{
c_x c_z & s_x c_z & s_z \cr
- c_x s_y s_z - s_x c_y e^{-i\delta} &
- s_x s_y s_z + c_x c_y e^{-i\delta} &
s_y c_z \cr 
- c_x c_y s_z + s_x s_y e^{-i\delta} & 
- s_x c_y s_z - c_x s_y e^{-i\delta} & 
c_y c_z \cr } \right ) 
\left ( \matrix{
e^{i\rho} & 0 & \cr
0 & e^{i\sigma} & 0 \cr
0 & 0 & 1 \cr} \right ) \; 
\end{equation}
with $s_x \equiv \sin\theta_x$, $c_x \equiv \cos\theta_x$ and so on.
It is obvious that $M$ depends on nine real parameters. Allowing one 
entry of $M$ to vanish, i.e., $M_{\alpha\beta} =0$, we obtain
\begin{equation}
\sum_{i=1}^3 \left (m_i V_{\alpha i} V_{\beta i} \right ) \; =\; 0 \; ,
\end{equation}
where $\alpha$ and $\beta$ run over $e$, $\mu$ and $\tau$. There
are totally six possible one-zero textures of $M$, illustrated in
Table 1 as patterns A, B, C, D, E and F. Pattern A and its 
consequences have somehow been discussed in Ref. \cite{Xing03}.

Now we consider whether one of three neutrino masses can vanish. Note
that the solar and atmospheric neutrino oscillations probe the 
mass-squared differences
\begin{eqnarray}
\Delta m^2_{\rm sun} & \equiv & \left | m^2_2 - m^2_1 \right | \; ,
\nonumber \\
\Delta m^2_{\rm atm} & \equiv & \left | m^2_3 - m^2_2 \right | \; ,
\end{eqnarray}
respectively. As the present solar neutrino data support 
$m_1 < m_2$ \cite{SNO,KM}, $m_2$ must not vanish. Thus we are left
with two possibilities \cite{Branco}: either $m_1 =0$ (normal 
hierarchy) or $m_3 =0$ (inverted hierarchy).

(1) If $m_1 =0$ holds, Eq. (3) can be simplified to
\begin{equation}
\xi \equiv \frac{m_2}{m_3} = 
\frac{|V_{\alpha 3} V_{\beta 3}|}{|V_{\alpha 2} V_{\beta 2}|} \;\; ,
\end{equation}
and
\begin{equation}
\arg \left (\frac{V_{\alpha 3} V_{\beta 3}}
{V_{\alpha 2} V_{\beta 2}} \right ) = \pm\pi \; .
\end{equation}
Eq. (6) implies a strong constraint on the CP-violating 
phases hidden in the matrix elements of $V$.
The ratio of $\Delta m^2_{\rm sun}$ and $\Delta m^2_{\rm atm}$
can be calculated by use of Eqs. (4) and (5). We obtain
\begin{equation}
R_\xi \; \equiv \; \frac{\Delta m^2_{\rm sun}}{\Delta m^2_{\rm atm}} 
\; =\; \frac{\xi^2}{\left | 1 - \xi^2 \right |} \;\; .
\end{equation}
As $R_\xi \sim 10^{-2}$ is required by current experimental data,
$\xi^2 \sim 10^{-2}$ must hold. Hence one arrives at a normal neutrino 
mass hierarchy. 

(2) If $m_3 =0$ holds, Eq. (3) can be simplified to
\begin{equation}
\zeta \equiv \frac{m_1}{m_2} = 
\frac{|V_{\alpha 2} V_{\beta 2}|}{|V_{\alpha 1} V_{\beta 1}|} \;\; ,
\end{equation}
and
\begin{equation}
\arg \left (\frac{V_{\alpha 2} V_{\beta 2}}
{V_{\alpha 1} V_{\beta 1}} \right ) = \pm\pi \; .
\end{equation}
Again, Eq. (9) implies a strong constraint on the CP-violating 
phases hidden in the matrix elements of $V$.
The ratio of $\Delta m^2_{\rm sun}$ and $\Delta m^2_{\rm atm}$
in this case is found to be
\begin{equation}
R_\zeta \; \equiv \; \frac{\Delta m^2_{\rm sun}}{\Delta m^2_{\rm atm}} 
\; =\; 1 ~ - ~ \zeta^2 \; .
\end{equation}
Because both $\zeta < 1$ and $R_\zeta \sim 10^{-2}$ are required,
only a very narrow range of $\zeta$ (even smaller than the range 
$0.95 < \zeta <1$) can satisfy Eq. (10).  

The explicit expressions of $\xi$ and $\zeta$ for each one-zero
texture of the neutrino mass matrix $M$ are given in 
Table 1 in terms of the flavor mixing
parameters $\theta_x$, $\theta_y$, $\theta_z$ and $\delta$.
In view of the recent SK \cite{SK}, SNO \cite{SNO}, 
KamLAND \cite{KM}, K2K \cite{K2K} and CHOOZ \cite{CHOOZ}
data on neutrino oscillations, we have 
$\Delta m^2_{\rm sun} \in [5.9, ~ 8.8] \times 10^{-5} ~ {\rm eV}^2$,
$\sin^2 \theta_x \in [0.25, ~ 0.40]$ \cite{Smirnov};
$\Delta m^2_{\rm atm} \in [1.65, ~ 3.25] \times 10^{-3} ~ {\rm eV}^2$,
$\sin^2 2\theta_y \in [0.88, ~ 1.00]$ \cite{Fogli}; and
$\sin^2 2\theta_z < 0.2$ at the $90\%$ confidence level.
Then we arrive at
$R_\xi = R_\zeta \in [0.018, 0.053]$ as well as
$\theta_x \in [30.0^\circ, 39.2^\circ]$,
$\theta_y \in [34.9^\circ, 55.1^\circ]$ and
$\theta_z \in [0.0^\circ, 13.3^\circ)$. These experimental results
allow us to examine which pattern of $M$ is phenomenologically
favored, although the Dirac phase $\delta$ is entirely unrestricted.

Taking arbitrary values of $\delta \in [0, 2\pi)$, we find that
patterns A, B and C with $m_1 =0$, in which the magnitudes of $R_\xi$
are suppressed by small $s^2_z$, can be compatible with current
neutrino oscillation data. Patterns D, E and F with $m_1 =0$ must
be discarded, because their predictions for $R_\xi$ are too big to
agree with the present experimental data. In the $m_3 =0$ case,
patterns B, C, D and F are allowed by current data; and patterns A 
and E are ruled out. Our 
numerical analyses show that pattern A with $m_3 =0$ yields
$R_\zeta \sim 0.5$ and pattern E with $m_3 =0$ yields $\zeta > 1$,
thus both patterns must be rejected. We are then left with seven
phenomenologically acceptable patterns of the neutrino mass matrix 
with one texture zero and one vanishing eigenvalue. Subsequently
we discuss each of the seven patterns in some detail.

\underline{Pattern A with $m_1 =0$:} ~ 
We have $\xi \propto s^2_z$ and $R_\xi \propto s^4_z$, whose exact
expressions are independent of the Dirac phase $\delta$. Considering
Eqs. (2) and (6) for $\alpha =\beta =e$, we obtain 
$\sigma =\pm \pi/2$. The other Majorana phase $\rho$ is completely
unrestricted. Because of $M_{ee} =0$ in this pattern, the effective 
mass of the neutrinoless double beta decay 
$\langle m\rangle_{ee} \equiv |M_{ee}|$ vanishes. This result has no
conflict with the present experimental data on 
$\langle m\rangle_{ee}$ \cite{HM}, namely, 
$\langle m\rangle_{ee} < 0.35$ eV
at the $90\%$ confidence level. It is remarkable
that three neutrino masses perform a clear hierarchy: 
$m_1 =0$, $m_2 = \sqrt{\Delta m^2_{\rm sun}} \sim 8\times 10^{-3}$ eV
and $m_3 \approx \sqrt{\Delta m^2_{\rm atm}} \sim 5\times 10^{-2}$ eV.

\underline{Pattern B with $m_1 =0$:} ~ 
We have $\xi \propto s_z$ and $R_\xi \propto s^2_z$, whose exact
expressions are weakly dependent on the Dirac phase $\delta$. 
Considering Eqs. (2) and (6) for $\alpha = e$ and $\beta =\mu$, we obtain 
$\sigma =(\Delta_1 \pm \pi)/2$, where
\begin{equation}
\tan \Delta_1 \; =\; \frac{c_x c_y s_\delta}{c_x c_y c_\delta -
s_x s_y s_z} \;\; ,
\end{equation}
with $s_\delta \equiv \sin\delta$ and $c_\delta \equiv \cos\delta$.
The other Majorana phase $\rho$ is entirely unrestricted. The neutrino
mass spectrum in this pattern is essentially the same as that in
pattern A with $m_1 =0$. Because $\langle m\rangle_{ee} \propto m_3 s_z$ 
holds, the effective mass of the neutrinoless double beta decay is
maximally of ${\cal O}(10^{-2})$ eV.

\underline{Pattern C with $m_1 =0$:} ~ 
We have $\xi \propto s_z$ and $R_\xi \propto s^2_z$, whose exact
expressions depend weakly on the Dirac phase $\delta$. 
Considering Eqs. (2) and (6) for $\alpha = e$ and $\beta =\tau$, we obtain 
$\sigma =(\Delta'_1 \pm \pi)/2$, where
\begin{equation}
\tan \Delta'_1 \; =\; \frac{c_x s_y s_\delta}{c_x s_y c_\delta +
s_x c_y s_z} \;\; . 
\end{equation}
One can see that $\Delta'_1 = \Delta_1$ will hold, if the replacement
$\theta_y \Longrightarrow \theta_y +\pi/2$ is made for Eq. (12).
Again, the Majorana phase $\rho$ is entirely unrestricted. The neutrino
mass spectrum in this pattern is essentially the same as that in
pattern A or pattern B with $m_1 =0$. As a result of 
$\langle m\rangle_{ee} \propto m_3 s_z$, the effective mass of the 
neutrinoless double beta decay is maximally of ${\cal O}(10^{-2})$ eV.

\underline{Pattern B with $m_3 =0$:} ~ If $\zeta$ and $R_\zeta$ are
expanded in powers of $s_z$, one can obtain $(1-\zeta) \propto s_z$ and 
$R_\zeta \propto s_z$, whose magnitudes depend strongly on the Dirac 
phase $\delta$. The explicit dependence of $R_\zeta$ on $\delta$ is
shown in Fig. 1, where current neutrino oscillation data have been
taken into account. Considering Eqs. (2) and (9) for $\alpha = e$ and 
$\beta =\mu$, we obtain $(\rho-\sigma) =(\Delta_2 \pm \pi)/2$, where
\begin{equation}
\tan \Delta_2 \; =\; \frac{s_y c_y s_z s_\delta}{s_x c_x 
(s^2_y s^2_z - c^2_y) + (s^2_x - c^2_x) s_y c_y s_z c_\delta} \;\; .
\end{equation}
Because of $m_3 =0$ and
$m_1 \approx m_2 = \sqrt{\Delta m^2_{\rm atm}} \sim 5\times 10^{-2}$ eV,
the mass spectrum of three neutrinos performs an inverted hierarchy.
In this case, the effective mass of the neutrinoless double beta decay 
$\langle m\rangle_{ee}$ is expected to be of ${\cal O}(10^{-2})$ eV or 
smaller.

\underline{Pattern C with $m_3 =0$:} ~ If $\zeta$ and $R_\zeta$ are
expanded in powers of $s_z$, one can also obtain $(1-\zeta) \propto s_z$ 
and $R_\zeta \propto s_z$, whose magnitudes depend strongly on the Dirac 
phase $\delta$. The explicit dependence of $R_\zeta$ on $\delta$ is
illustrated in Fig. 1, where the present experimental data have been
taken into account. Considering Eqs. (2) and (9) for $\alpha = e$ and 
$\beta =\tau$, we obtain $(\rho-\sigma) =(\Delta'_2 \pm \pi)/2$, where
\begin{equation}
\tan \Delta'_2 \; =\; \frac{s_y c_y s_z s_\delta}{s_x c_x 
(s^2_y - c^2_y s^2_z) + (s^2_x - c^2_x) s_y c_y s_z c_\delta} \;\; .
\end{equation}
It is easy to check that $\Delta'_2 = \Delta_2$ will hold, if the
replacement $\theta_y \Longrightarrow \theta_y + \pi/2$ is made.
The neutrino mass spectrum in this pattern is essentially the same
as that in pattern B with $m_3 =0$. Thus we expect that 
$\langle m\rangle_{ee}$ is of ${\cal O}(10^{-2})$ eV or smaller.

\underline{Pattern D with $m_3 =0$:} ~ If $\zeta$ is expanded in powers 
of $s_z$, one can obtain $\zeta \propto c^2_x/s^2_x$. Then the smallness
and positiveness of $R_\zeta$ depends on the Dirac phase $\delta$ in an 
indirect way. The numerical dependence of $R_\zeta$ on $\delta$ is
shown in Fig. 2, where current neutrino oscillation data have been
taken into account. Considering Eqs. (2) and (9) for 
$\alpha = \beta =\mu$, we get $(\rho-\sigma) =\Delta_2 \pm \pi/2$, 
where $\Delta_2$ has been given in Eq. (13). We also have
$m_1 \approx m_2 = \sqrt{\Delta m^2_{\rm atm}} \sim 5\times 10^{-2}$ eV
in this pattern. Thus the neutrino mass spectrum is essentially the same 
as that in pattern B or C with $m_3 =0$. Again, the size of 
$\langle m\rangle_{ee}$ is anticipated to be of ${\cal O}(10^{-2})$ eV 
or smaller.

\underline{Pattern F with $m_3 =0$:} ~ If $\zeta$ is expanded in powers 
of $s_z$, one can obtain $\zeta \propto c^2_x/s^2_x$. The smallness
and positiveness of $R_\zeta$ depends on the Dirac phase $\delta$ in an 
indirect way. We illustrate the numerical dependence of $R_\zeta$ on 
$\delta$ in Fig. 2. Considering Eqs. (2) and (9) for 
$\alpha = \beta =\tau$, we get $(\rho-\sigma) =\Delta'_2 \pm \pi/2$, 
where $\Delta'_2$ has been given in Eq. (14). It is found that
the neutrino mass spectrum in this pattern is essentially the same 
as that in pattern B, C or D with $m_3 =0$. Thus we anticipate 
$\langle m\rangle_{ee}$ to be of ${\cal O}(10^{-2})$ eV or smaller.

In summary, we have examined whether the Majorana neutrino mass matrix 
$M$ with one texture zero and one vanishing eigenvalue can interpret
the experimental data on solar, atmospheric and reactor neutrino
oscillations. We find that three one-zero patterns of $M$ with 
$m_1 =0$ (normal hierarchy) and four one-zero patterns of $M$ with
$m_3 =0$ are compatible with current data. Thus these simple
textures of $M$ are interesting and useful for model building, in order 
to reach some deeper understanding of the origin of lepton masses and 
flavor mixing. 

It is worth mentioning that the one-zero pattern F with $m_3 =0$
has recently been derived by He \cite{He} from a modified version of
the Zee model \cite{Zee}. One may examine that other one-zero 
textures of $M$ with $m_1 =0$ or $m_3 =0$ are possible to result from 
similar modifications of the Zee model. If the seesaw mechanism \cite{SS}
is taken into account, one may also derive possible one-zero patterns of
the heavy right-handed Majorana neutrino mass matrix in a phenomenological
way with a few reasonable assumptions (see Ref. \cite{Hambye} for some 
detailed discussions).

Of course, a specific texture of the neutrino
mass matrix $M$ may not be preserved to all orders or at any energy 
scales in the unspecified interactions from which neutrino masses are
generated \cite{Review}. Whether $m_1 =0$ or $m_3 =0$ is stable against 
radiative corrections should also be taken into account in a concrete 
model, into which the simple ans$\rm\ddot{a}$tze of $M$ under discussion 
can be incorporated. Between the electroweak scale ($m^{~}_Z$) and the
mass scale of the lightest right-handed Majorana neutrino 
($m^{~}_{\rm R1}$) in the seesaw mechanism \cite{SS}, however, the 
running behavior of the effective left-handed neutrino mass matrix $M$ 
is rather simple at the one-loop level (see, e.g., Ref. \cite{Ellis}):
\begin{equation}
M(m^{~}_{\rm R1}) \; =\; I^{-1}_g I^{-1}_t 
\left ( \matrix{
\sqrt{I_e} & 0 & 0 \cr
0 & \sqrt{I_\mu} & 0 \cr
0 & 0 & \sqrt{I_\tau} \cr} \right )^{-1} M(m^{~}_Z)
\left ( \matrix{
\sqrt{I_e} & 0 & 0 \cr
0 & \sqrt{I_\mu} & 0 \cr
0 & 0 & \sqrt{I_\tau} \cr} \right )^{-1} \; ,
\end{equation}
where $I_g$, $I_t$ and $I_\alpha$ (for $\alpha = e, \mu, \tau$) stand
respectively for the evolution functions of the gauge couplings, the
top-quark Yukawa coupling and the charged-lepton Yukawa coupling 
eigenvalues. It becomes obvious that the vanishing eigenvalue and
texture zeros of $M$ at $m^{~}_Z$ remain to hold at $m^{~}_{\rm R1}$,
at least at the one-loop level under consideration \cite{Ellis}. 
This interesting
observation implies that our classification of $M$ with texture zeros
or vanishing eigenvalues at low energy scales {\it does} make some 
sense for model building at a much higher energy scale. 

\vspace{0.5cm}

I like to thank X.J. Bi for useful discussions. This work was supported 
in part by the National Natural Science Foundation of China.

\newpage

\begin{table}
\caption{Six possible one-zero textures of the neutrino mass matrix 
$M$, in which all the non-vanishing entries are symbolized by $\times$'s.
The neutrino mass ratio $\xi \equiv m_2/m_3$ in the $m_1 =0$ case or
$\zeta \equiv m_1/m_2$ in the $m_3 =0$ case is expressed in terms of 
the flavor mixing parameters $\theta_x$, $\theta_y$, $\theta_z$ and 
$\delta$. For simplicity, $c_\delta \equiv \cos\delta$ has been used.}
\vspace{0.5cm}
\begin{center}
\begin{tabular}{lll} 
Pattern of $M$ & ~~~~& Case of $m_1 = 0$ (normal hierarchy) \\ \hline 
$\rm A:$ 
$\left ( \matrix{
{\bf 0} & \times & \times \cr
\times  & \times & \times \cr
\times  & \times & \times \cr} \right )$
&& 
$\displaystyle \xi = \frac{s^2_z}{s^2_x c^2_z}$
\\  
$\rm B:$ 
$\left ( \matrix{
\times  & {\bf 0} & \times \cr
{\bf 0} & \times & \times \cr
\times  & \times & \times \cr} \right )$
&& 
$\displaystyle \xi = \frac{s_y s_z}
{s_x \sqrt{c^2_x c^2_y - 2s_x c_x s_y c_y s_z c_\delta +
s^2_x s^2_y s^2_z}}$
\\ 
$\rm C:$ 
$\left ( \matrix{
\times  & \times & {\bf 0} \cr
\times  & \times & \times \cr
{\bf 0} & \times & \times \cr} \right )$
&& 
$\displaystyle \xi = \frac{c_y s_z}
{s_x \sqrt{c^2_x s^2_y + 2s_x c_x s_y c_y s_z c_\delta +
s^2_x c^2_y s^2_z}}$
\\ 
$\rm D:$ 
$\left ( \matrix{
\times & \times  & \times \cr
\times & {\bf 0} & \times \cr
\times & \times  & \times \cr} \right )$
&& 
$\displaystyle \xi = \frac{s^2_y c^2_z}
{c^2_x c^2_y - 2s_x c_x s_y c_y s_z c_\delta +
s^2_x s^2_y s^2_z}$
\\ 
$\rm E:$ 
$\left ( \matrix{
\times & \times  & \times  \cr
\times & \times  & {\bf 0} \cr
\times & {\bf 0} & \times  \cr} \right )$
&& 
$\displaystyle \xi = \frac{s_y c_y c^2_z}
{\sqrt{c^2_x c^2_y - 2s_x c_x s_y c_y s_z c_\delta +
s^2_x s^2_y s^2_z}
\sqrt{c^2_x s^2_y + 2s_x c_x s_y c_y s_z c_\delta +
s^2_x c^2_y s^2_z}}$
\\ 
$\rm F:$ 
$\left ( \matrix{
\times & \times & \times  \cr
\times & \times & \times  \cr
\times & \times & {\bf 0} \cr} \right )$
&& 
$\displaystyle \xi = \frac{c^2_y c^2_z}
{c^2_x s^2_y + 2s_x c_x s_y c_y s_z c_\delta +
s^2_x c^2_y s^2_z}$
\\ \hline\hline
Pattern of $M$ & ~~~~ & Case of $m_3 = 0$ (inverted hierarchy) \\ \hline 
$\rm A:$ 
$\left ( \matrix{
{\bf 0} & \times & \times \cr
\times  & \times & \times \cr
\times  & \times & \times \cr} \right )$
&& 
$\displaystyle \zeta = \frac{s^2_x}{c^2_x}$
\\  
$\rm B:$ 
$\left ( \matrix{
\times  & {\bf 0} & \times \cr
{\bf 0} & \times & \times \cr
\times  & \times & \times \cr} \right )$
&& 
$\displaystyle \zeta = \frac{
s_x \sqrt{c^2_x c^2_y - 2s_x c_x s_y c_y s_z c_\delta +
s^2_x s^2_y s^2_z}}
{c_x \sqrt{s^2_x c^2_y + 2s_x c_x s_y c_y s_z c_\delta +
c^2_x s^2_y s^2_z}}$
\\ 
$\rm C:$ 
$\left ( \matrix{
\times  & \times & {\bf 0} \cr
\times  & \times & \times \cr
{\bf 0} & \times & \times \cr} \right )$
&& 
$\displaystyle \zeta = \frac{
s_x \sqrt{c^2_x s^2_y + 2s_x c_x s_y c_y s_z c_\delta +
s^2_x c^2_y s^2_z}}
{c_x \sqrt{s^2_x s^2_y - 2s_x c_x s_y c_y s_z c_\delta +
c^2_x c^2_y s^2_z}}$
\\ 
$\rm D:$ 
$\left ( \matrix{
\times & \times  & \times \cr
\times & {\bf 0} & \times \cr
\times & \times  & \times \cr} \right )$
&& 
$\displaystyle \zeta = \frac{
c^2_x c^2_y - 2s_x c_x s_y c_y s_z c_\delta +
s^2_x s^2_y s^2_z}
{s^2_x c^2_y + 2s_x c_x s_y c_y s_z c_\delta +
c^2_x s^2_y s^2_z}$
\\ 
$\rm E:$ 
$\left ( \matrix{
\times & \times  & \times  \cr
\times & \times  & {\bf 0} \cr
\times & {\bf 0} & \times  \cr} \right )$
&& 
$\displaystyle \zeta = \frac{
\sqrt{c^2_x c^2_y - 2s_x c_x s_y c_y s_z c_\delta +
s^2_x s^2_y s^2_z}
\sqrt{c^2_x s^2_y + 2s_x c_x s_y c_y s_z c_\delta +
s^2_x c^2_y s^2_z}}
{\sqrt{s^2_x c^2_y + 2s_x c_x s_y c_y s_z c_\delta +
c^2_x s^2_y s^2_z}
\sqrt{s^2_x s^2_y - 2s_x c_x s_y c_y s_z c_\delta +
c^2_x c^2_y s^2_z}}$
\\ 
$\rm F:$ 
$\left ( \matrix{
\times & \times & \times  \cr
\times & \times & \times  \cr
\times & \times & {\bf 0} \cr} \right )$
&& 
$\displaystyle \zeta = \frac{
c^2_x s^2_y + 2s_x c_x s_y c_y s_z c_\delta +
s^2_x c^2_y s^2_z}
{s^2_x s^2_y - 2s_x c_x s_y c_y s_z c_\delta +
c^2_x c^2_y s^2_z}$
\\ 
\end{tabular}
\end{center}
\end{table}
\normalsize

\newpage

\begin{figure}[t]
\vspace{1.5cm}
\epsfig{file=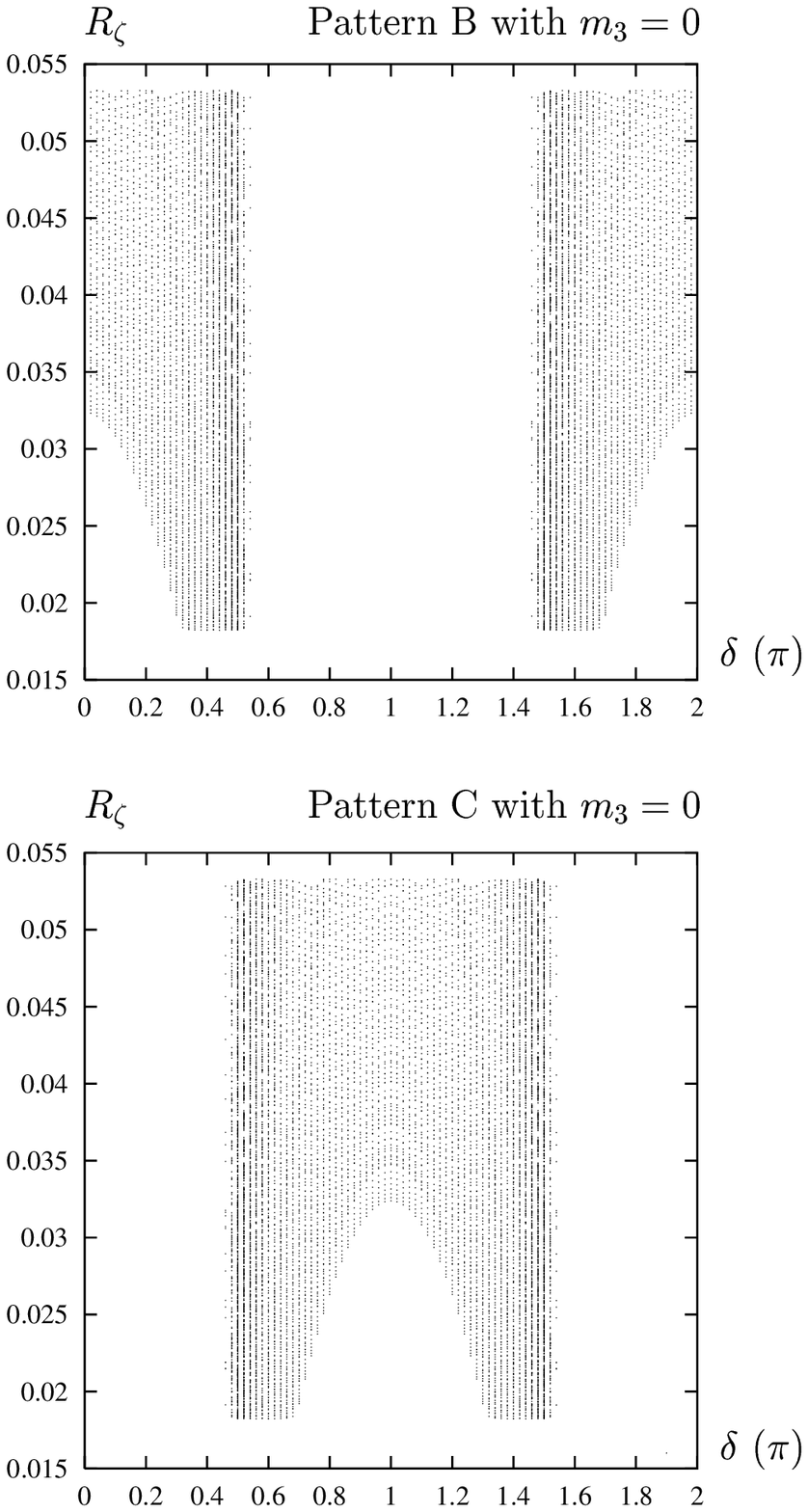,bbllx=1.8cm,bblly=5cm,bburx=18cm,bbury=26cm,%
width=13.5cm,height=18cm,angle=0,clip=}
\vspace{-0.3cm}
\caption{The numerical dependence of $R_\zeta$ on the Dirac phase 
$\delta$ for patterns B and C with $m_3 =0$, where current neutrino
oscillation data have been taken into account.}
\end{figure}

\newpage

\begin{figure}[t]
\vspace{3.5cm}
\epsfig{file=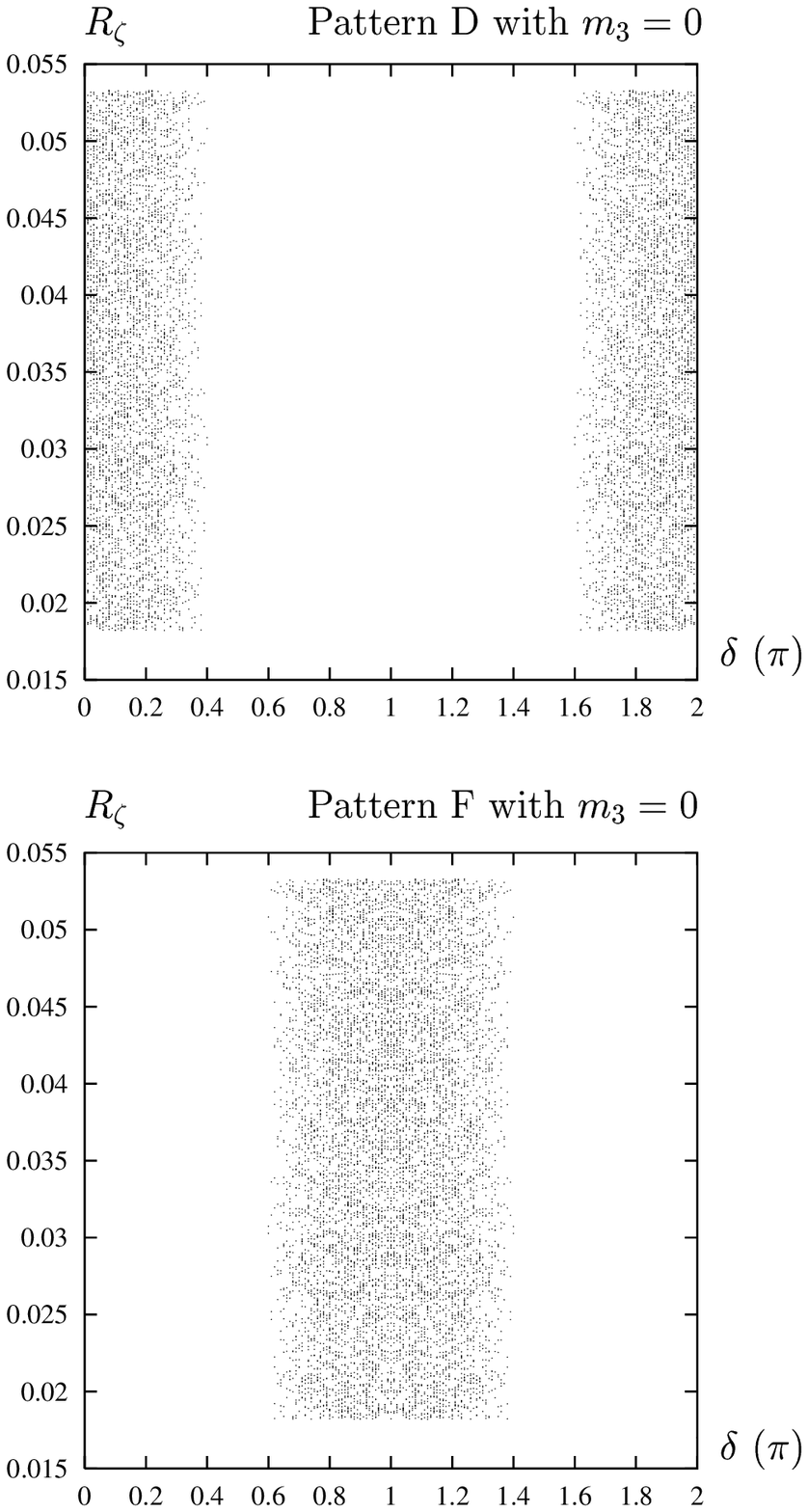,bbllx=1.8cm,bblly=5cm,bburx=18cm,bbury=26cm,%
width=13.5cm,height=18cm,angle=0,clip=}
\vspace{-0.3cm}
\caption{The numerical dependence of $R_\zeta$ on the Dirac phase 
$\delta$ for patterns D and F with $m_3 =0$, where current neutrino
oscillation data have been taken into account.}
\end{figure}


\newpage
\begin{thebibliography}{99}
\bibitem{SK} For a review, see: C.K. Jung, C. McGrew, T. Kajita,
and T. Mann, Ann. Rev. Nucl. Part. Sci. {\bf 51}, 451 (2001).

\bibitem{SNO} SNO Collaboration, Q.R. Ahmad {\it et al.},
Phys. Rev. Lett. {\bf 89}, 011301 (2002); Phys. Rev. Lett.
{\bf 89}, 011302 (2002).

\bibitem{KM} KamLAND Collaboration, K. Eguchi {\it et al.},
Phys. Rev. Lett. {\bf 90}, 021802 (2003).

\bibitem{K2K} K2K Collaboration, M.H. Ahn {\it et al.},
Phys. Rev. Lett. {\bf 90}, 041801 (2003).

\bibitem{FGM} P.H. Frampton, S.L. Glashow, and D. Marfatia,
Phys. Lett. B {\bf 536}, 79 (2002);
Z.Z. Xing, Phys. Lett. B {\bf 530}, 159 (2002);
Phys. Lett. B {\bf 539}, 85 (2002);
A. Kageyama, S. Kaneko, N. Shimoyama, and M. Tanimoto,
Phys. Lett. B {\bf 538}, 96 (2002);
W.L. Guo and Z.Z. Xing, Phys. Rev. D {\bf 67}, 053002 (2003);
M. Honda, S. Kaneko, and M. Tanimoto, hep-ph/0303227.

\bibitem{Branco} G.C. Branco, R. Gonz$\rm\acute{a}$lez, F.R. Joaquim,
and T. Yanagida, Phys. Lett. B {\bf 562}, 265 (2003).

\bibitem{FX01} H. Fritzsch and Z.Z. Xing, 
Phys. Lett. B {\bf 517}, 363 (2001);
Z.Z. Xing, Phys. Rev. D {\bf 64}, 073014 (2001);
Phys. Rev. D {\bf 65}, 113010 (2002).

\bibitem{Xing03} Z.Z. Xing, hep-ph/0305195.

\bibitem{CHOOZ} CHOOZ Collaboration, M. Apollonio {\it et al.},
Phys. Lett. B {\bf 420}, 397 (1998);
Palo Verde Collaboration, F. Boehm {\it et al.},
Phys. Rev. Lett. {\bf 84}, 3764 (2000).

\bibitem{Smirnov} P.C. de Holanda and A.Yu. Smirnov, hep-ph/0212270;
and references therein.

\bibitem{Fogli} G.L. Fogli, E. Lisi, A. Marrone, and D. Montanino,
hep-ph/0303064.

\bibitem{HM} Heidelberg-Moscow Collaboration, H.V. Klapdor-Kleingrothaus,
hep-ph/0103074; C.E. Aalseth {\it et al.}, hep-ex/0202026;
and references cited therein.

\bibitem{He} X.G. He, hep-ph/0307172.

\bibitem{Zee} A. Zee, Phys. Lett. B {\bf 93}, 389 (1980);
Phys. Lett. B {\bf 161}, 141 (1985);
L. Wolfenstein, Nucl. Phys. B {\bf 175}, 93 (1980).

\bibitem{SS} T. Yanagida, in {\it Proceedings of the Workshop on 
Unified Theory and the Baryon Number of the Universe}, edited by 
O. Sawada and A. Sugamoto (KEK, 1979); M. Gell-Mann, P. Ramond, and
R. Slansky, in {\it Supergravity}, edited by F. van Nieuwenhuizen
and D. Freedman (North Holland, Amsterdam, 1979); R.N. Mohapatra and
G. Senjanovic, Phys. Rev. Lett. {\bf 44}, 912 (1980).

\bibitem{Hambye} R. Barbieri, T. Hambye, and A. Romanino,
JHEP {\bf 0303}, 017 (2003).

\bibitem{Review} For recent reviews with extensive references, see, e.g.,
H. Fritzsch and Z.Z. Xing, Prog. Part. Nucl. Phys. {\bf 45}, 1 (2000);
G. Altarelli and F. Feruglio, hep-ph/0306265.

\bibitem{Ellis} J. Ellis and S. Lola, Phys. Lett. B {\bf 458}, 310 (1999);
H. Fritzsch and Z.Z. Xing, in Ref. \cite{Review}; and references therein.
\end{thebibliography}
\end{document}